\documentclass{mem}
\usepackage{natbib}\usepackage{txfonts}\usepackage{balance}
\usepackage{graphicx}
\usepackage{txfonts}
\usepackage[a4paper]{hyperref}
\idline{79}{1}

\usepackage{ulem}
\begin{document}

\title{Fundamental properties of low-mass stars}

   \subtitle{}

\author{
I. Ribas\inst{1,2},
J.~C. Morales\inst{2}, 
C. Jordi\inst{2,3},
I. Baraffe\inst{4},
G. Chabrier\inst{4},
\and J. Gallardo\inst{4}
          }

\offprints{I. Ribas}

\institute{
Institut de Ci\`encies de l'Espai (CSIC), Campus UAB, Facultat de Ci\`encies, Torre C5-parell-2a planta, 08193 Bellaterra, Spain
\email{iribas@ieec.uab.es}
\and
Institut d'Estudis Espacials de Catalunya (IEEC), Edif. Nexus, C/Gran Capit\`a, 2-4, 08034 Barcelona, Spain
\and
Departament d'Astronomia i Meteorologia, Universitat de Barcelona, Avda. Diagonal 647, 08028 Barcelona, Spain
\and 
\'Ecole Normale Sup\'erieure de Lyon, CRAL (UMR CNRS 5574), Universit\'e de Lyon, France
}

\authorrunning{Ribas et al.}

\titlerunning{Fundamental properties of low-mass stars}

\abstract{
Numerous investigations on the fundamental properties of low-mass stars 
using eclipsing binaries indicate a strong discrepancy between theory and 
observations that is still awaiting explanation. Current models seem to 
predict radii for stars less massive than the Sun that are some 10\% 
smaller than observed, while their effective temperatures are some 5\% 
larger. Here we discuss recent new observational data that are relevant to 
this issue and review the progress made in understanding the origin of the 
important differences with theoretical calculations. Notably, we provide 
evidence that stellar activity may be responsible for the mismatch between 
observations and theory through two different channels: inhibition of 
convection or effects of a significant starspot coverage. The activity 
hypothesis is put to a test with observational diagnostics and some of the 
consequences of the large starspot coverage are evaluated. We conclude 
that stellar activity likely plays a key role in defining the properties 
of active low-mass stars and that this should be properly taken into 
account when investigating young, active stars in clusters or star-forming 
regions.

}
\maketitle{}

\section{Introduction}
 
Evidence collected over the past years convincingly shows that the 
components of low-mass eclipsing binary stars are not adequately 
reproduced by current evolution models. The observations yield stars that 
are systematically larger and cooler than theoretical calculations by 
about 10\% and 5\%, respectively. But, in contrast, these binary 
components are found to have luminosities that are in good agreement with 
those of single stars and also with model predictions. This is the reason 
why studies focused solely on the mass-luminosity relationship did not 
reveal any mismatch between observations and theory and, further, obtained 
very tight mass and luminosity correspondence when using infrared 
photometry (e.g., \citealt{Delfosse2000}).

\begin{figure*}[t] 
\centering 
\includegraphics[scale=0.4]{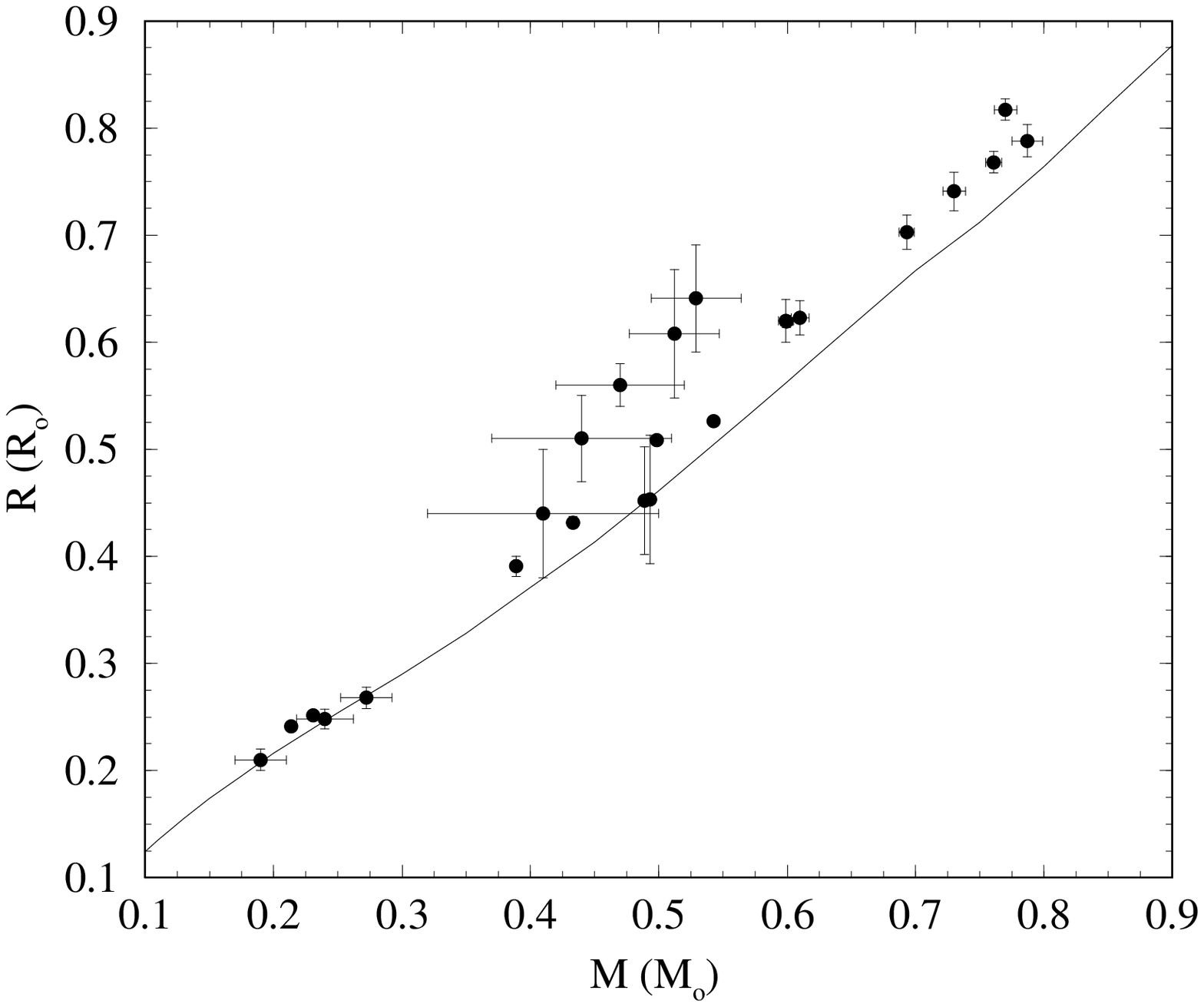} 
\includegraphics[scale=0.4]{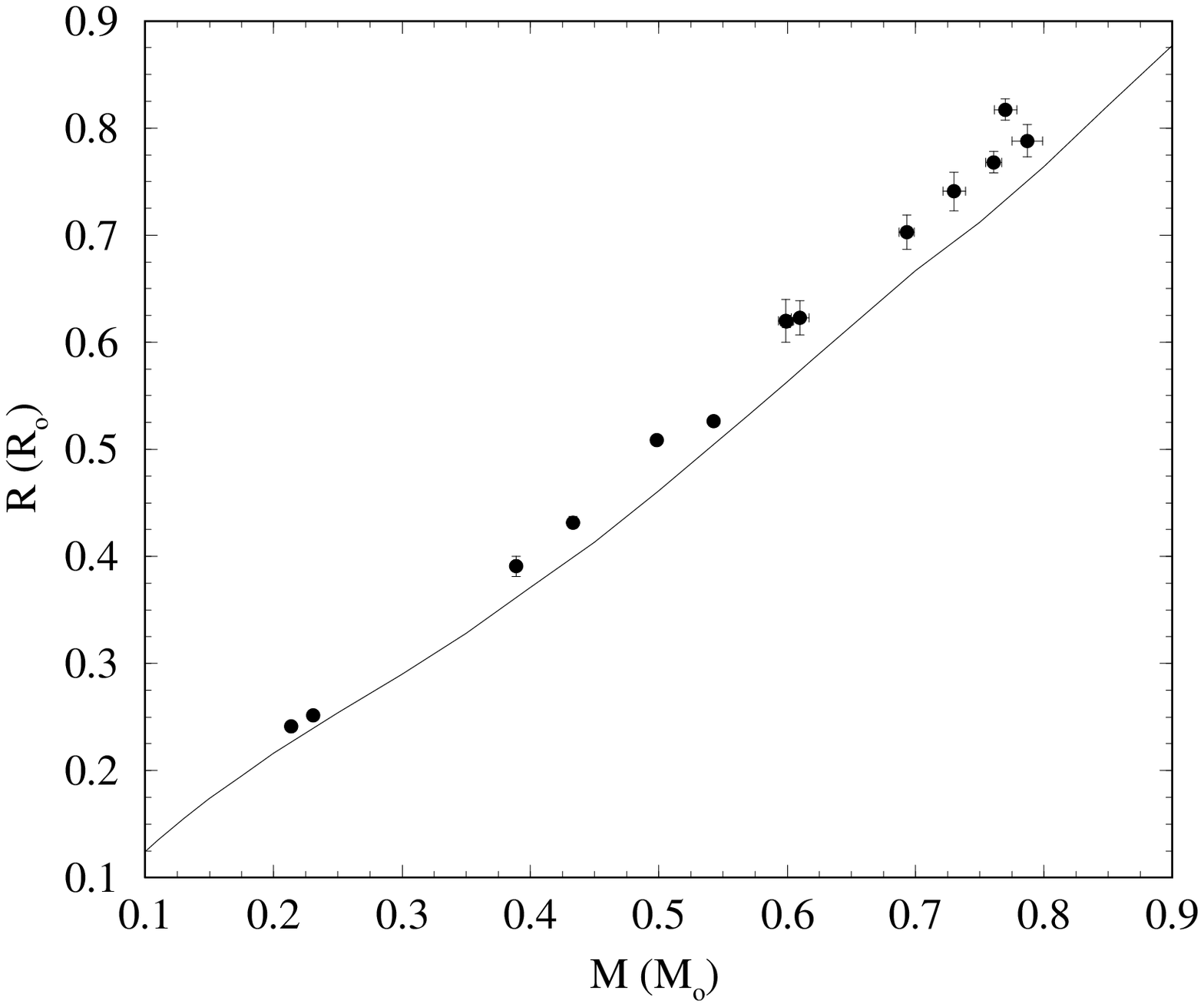} 
\caption{\footnotesize 
Left: $M$--$R$ plot for low-mass eclipsing binary stars with empirical 
determinations. Right: Same but for stars with uncertainties below 3\%. 
The solid line represents a theoretical isochrone of 1 Gyr calculated with 
the \citet{Baraffe1998} models.} 
\label{fig:fig_EBs} 
\end{figure*}

A revision of the available observational evidence and some hints about 
the possible resolutions of the discrepancies were provided by 
\citet{Ribas2006}. We refer to that study for background information and 
detailed references. Here we shall only discuss new results appeared since 
2006 and also the progress made in understanding the origin of the 
difference between model predictions and observations of fundamental 
properties of low-mass stars. 

\section{Recent observational data}

New low-mass eclipsing binaries have resulted in the past two years from 
dedicated monitoring of carefully selected targets from photometric 
databases of variability surveys. These have been published by 
\citet{LopezMorales-Shaw2006} and \citet{LopezMorales2006}. In addition, 
the low-mass eclipsing binary reported by \citet{Hebb2006} resulted from a 
deep targeted search in several open clusters. In other cases, low-mass 
eclipsing binaries were serendipitously discovered over the course of 
photometric monitoring campaigns with different scientific aims, and later 
analyzed specifically in detail, such as the objects studied by 
\citet{Bayless2006}, \citet{Young2006}, and \citet{Blake2007}. A 
mass-radius plot of all presently known low-mass stars in detached 
eclipsing binaries is provided in Fig. \ref{fig:fig_EBs} (left), while 
Fig. \ref{fig:fig_EBs} (right) shows only those objects that have reported 
error bars in both masses and radii below 3\%. The systematic offset of 
5--10\% between the observations and the 1 Gyr isochrone from the models 
of \citet{Baraffe1998} is apparent.

\begin{figure}[t!]
\centering
\resizebox{\hsize}{!}{\includegraphics[clip=true]{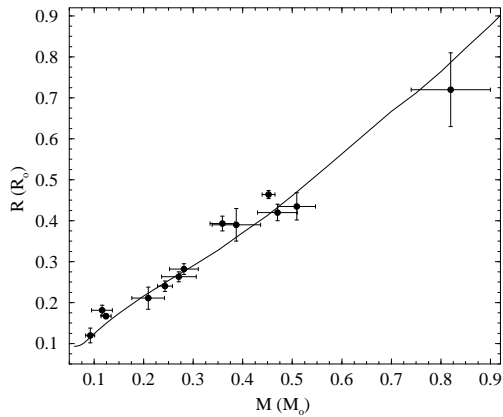}}
\caption{\footnotesize
Same as Fig. \ref{fig:fig_EBs} (left) but for single-line and 
single-eclipse systems.}
\label{fig:fig_SLSE}
\end{figure}

In addition to the ``classical'' eclipsing binaries, there has been an 
increasing number of discoveries resulting from follow-up of planetary 
transit candidates. In some instances, the object responsible for the 
transit was found not to be a planet but an M-dwarf secondary to a 
F-G-type primary star. This is the case of the recent study by 
\citet{Beatty2007} of one of the HAT network planetary candidates. These 
objects are single-line and single-eclipse binaries that directly provide 
a value of the stellar density but that require certain assumptions (e.g., 
orbital synchronization) to derive the actual masses and radii. A similar 
technique exploited by \citet{Torres2007} has provided masses and radii 
(dependent on the assumed effective temperature) for the M-type star 
GJ~436, which hosts a transiting exoplanet. A mass-radius plot of all the 
objects resulting from single-line radial velocities and single-eclipse 
light curves is shown in Fig. \ref{fig:fig_SLSE}. The only object in this 
sample with an error bar below 3\% (making it a reliable test of models) 
is GJ~436. Interestingly, it also shows the 10\% radius differential with 
model predictions \citep{Torres2007}.

Finally, there is an additional source of fundamental properties of 
low-mass stars from direct interferometric measurements. This was the 
method applied in the recent study of \citet{Berger2006}, where the 
authors report measurements for six M dwarfs with the CHARA array and 
suggested a correlation between radius differences and stellar 
metallicity.

In this context it is also worth reviewing the progress made in 
understanding the properties of the only known eclipsing binary with brown 
dwarf components. Following the discovery analysis of \citet{Stassun2006}, 
the recent analysis of \citet{Stassun2007} has further refined the 
properties of the binary components and strengthened the case for a 
reversed temperature ratio, which is difficult to reconcile with model 
calculations.

\section{Stellar activity hypothesis: a theoretical framework}

As discussed by \citet{Ribas2006} and \citet{Torres2006} there is a 
property that distinguishes low-mass stars in close binaries from those 
that are single objects, and this is the presence of tidal interactions 
that force the component stars to rotate in orbital synchrony. Because of 
strong observational biases, all low-mass eclipsing binaries found so far 
have periods well below 10 days, and, thus, rotation periods also below 10 
days. Using the analysis of, e.g., \citet{Pizzolato2003} it is easy to 
realize that M stars with such short rotation periods will experience high 
levels of magnetic activity. This is also confirmed by the fact that most 
low-mass eclipsing binaries are strong X-ray sources and also they show 
intense emission in the H$\alpha$ Balmer line.

From these observational facts it is sensible to consider a hypothesis by 
which the differences between model predictions and eclipsing binary 
observations arise from the high levels of magnetic activity of the 
component stars. Careful consideration of the effects of magnetic fields 
in low-mass stellar evolution models was carried out by 
\citet{Mullan2001}, using some prescriptions from \citet{Gough1966}. The 
authors pointed out that the inclusion of magnetic fields potentially has 
a moderate impact on the overall stellar properties. More recently, 
\citet{Chabrier2007} have performed a more realistic treatment targeted 
specifically to resolving the current differences between observations and 
models. The authors have considered two scenarios. Firstly, a scenario 
that considers the effect of magnetic fields in causing inhibition of the 
convective energy transport. And secondly, a scenario similar to that 
proposed by \citet{LopezMorales2005} and based on simple energy 
conservation arguments in a spot-covered surface.

The study of \citet{Chabrier2007} concludes that both mechanisms alter the 
properties of the star sufficiently to explain the observed radius and 
temperature discrepancies. In the case of the inhibition of convection, 
this was tested by setting the mixing-length parameter to lower values 
than that yielded by the standard solar model. The result on the structure 
of the star is that it shows a higher radius and slightly lower effective 
temperature. Tests show that good agreement with the observations is 
obtained for a mixing-length parameter of 0.5 pressure scale heights in 
the case of the more massive stars of the sample (i.e., 
$M>0.6$~M$_{\odot}$). However, less massive stars are little affected by 
the lower mixing-length parameter and agreement with observations is only 
obtained when decreasing to (possibly unphysical) values of 0.1 pressure 
scale heights or less. This is well illustrated in figure 1 of 
\citet{Chabrier2007}.

\begin{figure*}[t!]
\centering
\includegraphics[scale=0.39]{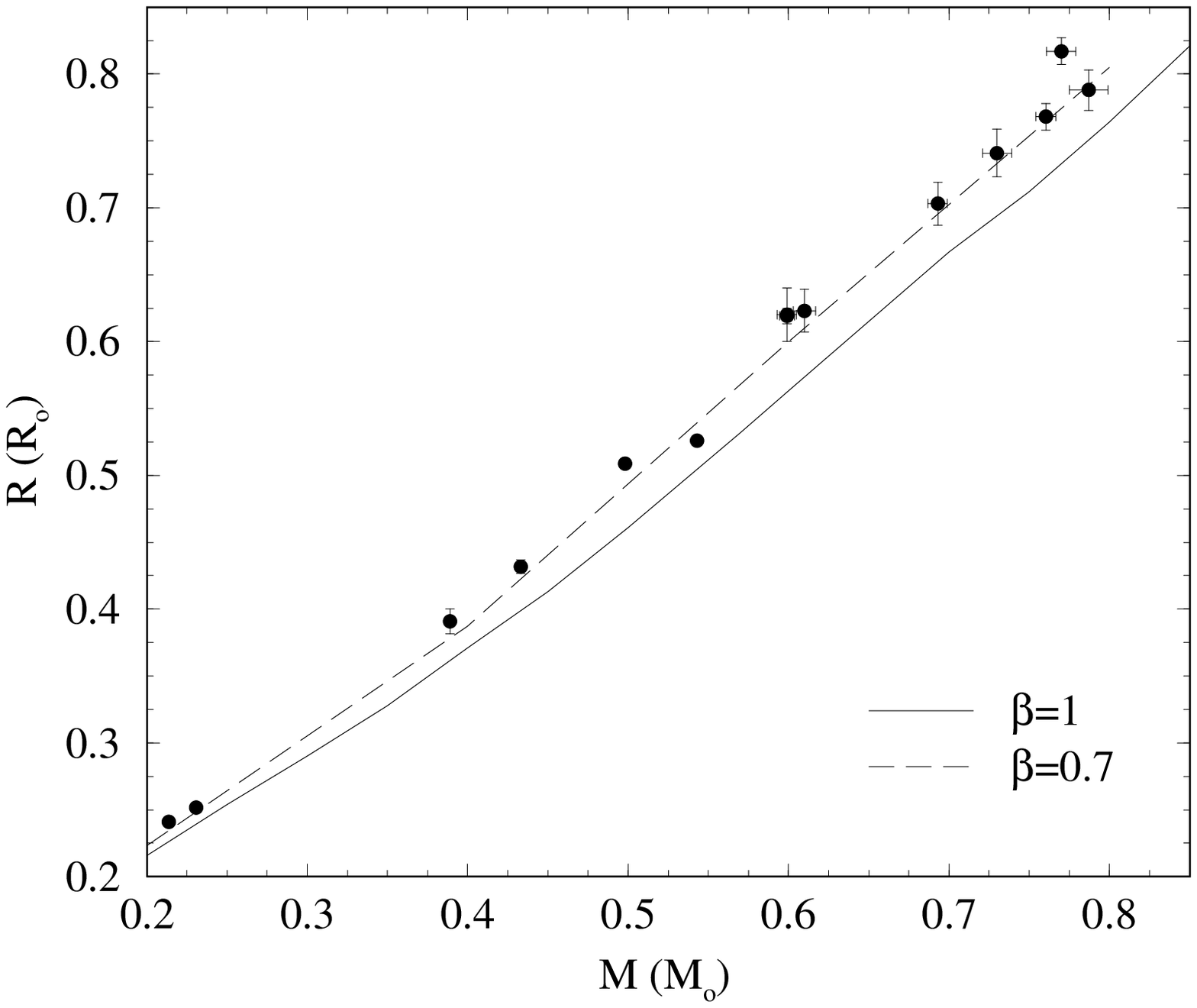}
\includegraphics[scale=0.39]{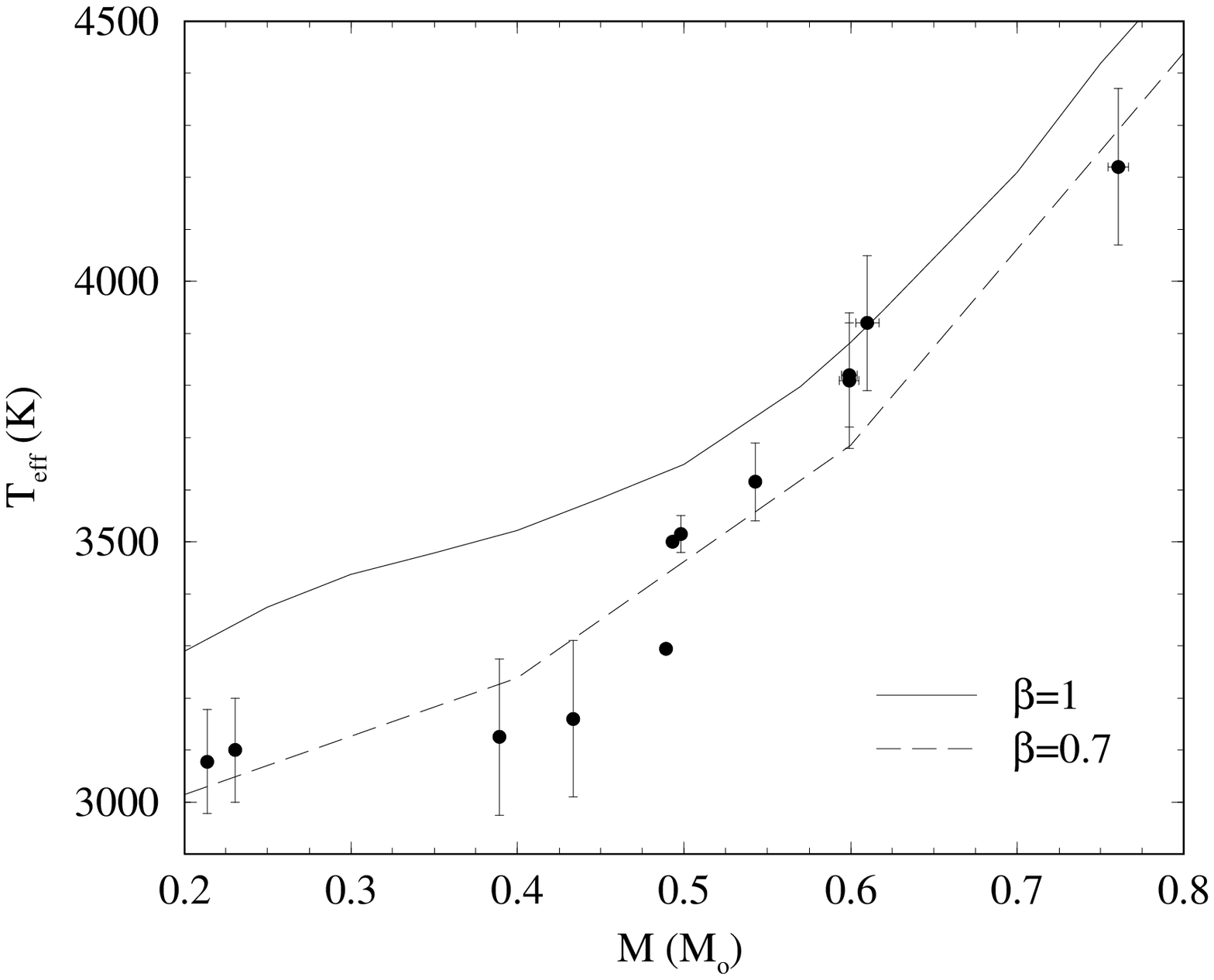}
\caption{\footnotesize
Comparison of the best known eclipsing binary systems with models using 
$\beta$ values of 0 and 0.3. Left: $M$--$R$ plot. Right: $M$--$T_{\rm 
eff}$ plot.}
\label{fig:fig_models}
\end{figure*}

In the case of the direct effect of starspot coverage, the analysis can be 
carried out by assuming a new luminosity $L'$ expressed as
\begin{equation}
L'= (1-\beta) \, 4\pi \, R'^2 \, \sigma {T_{\rm eff}'}^4
\end{equation}
where $R'$ is the modified radius, $T_{\rm eff}'$ is the modified 
effective temperature and $\beta$ is the factor by which starspots block 
the outgoing luminosity because of their lower temperature. From the 
calculations of models with this modified boundary condition, 
\citet{Chabrier2007} found that the radii of low-mass stars in eclipsing 
binaries in the entire mass domain can be reproduced with a $\beta$ 
parameter of about 0.3. Assuming a spot temperature contrast of about 15\% 
(or 500~K), the results indicate that starspots cover approximately half 
of the stellar surface. In this scenario, the modified stellar effective 
temperatures are also found to agree with the observations as the total 
stellar luminosity is nearly invariant. In Fig. \ref{fig:fig_models} we 
show a mass-radius plot and a radius-effective temperature plot that 
illustrate the good agreement between observations and models when 
considering the effects of starspots.

Thus, although conclusive evidence on which of the two scenarios (or 
perhaps a combination) is the most reliable to explain the observations is 
still lacking, the results of the analysis strongly suggest that stellar 
activity is a key element in understanding the properties of low-mass 
stars such as their radii and effective temperatures.

\section{Stellar activity effects: test with observations}

In this section we consider various observational evidence in the context 
of the stellar activity hypothesis. One of the straightforward questions 
to address is the universality of the hypothesis. In other words, if 
activity is thought to be responsible for the observed radius and 
effective temperature discrepancy in eclipsing binaries, do single active 
stars also show the same effect? Indeed, active stars have long been 
recognized to define a slightly offset sequence in the color-magnitude 
diagram (e.g., \citealt{Stauffer1986}). The results from the eclipsing 
binaries and the theoretical studies discussed above allow for a fresh 
look at the problem. From the available evidence it can be assumed that 
the luminosities of low-mass stars are not significantly affected by 
stellar activity. The studies by \citet{MoralesCefalu} and 
\citet{Morales2007} show that a comparison of active and inactive stars of 
the same luminosity indeed reveals a systematic temperature offset. The 
effective temperature differential can be translated into a radius 
difference that is of the same order as that found in eclipsing binaries. 
This result generalizes the activity effects on stellar properties to any 
star, either single or binary. Full details on the analysis can be found 
in the references above.

The activity hypothesis was also investigated by \citet{LopezMorales2007}, 
who collected values of rotational velocities and, eventually, X-ray 
luminosities of a sample of low-mass eclipsing binaries. Then, the radius 
discrepancies were searched for correlations with such X-ray luminosities, 
which are proxies to the overall stellar activity. The author identified 
significant correlations in the binary sample but not so in single stars 
with interferometric radius measurements. However, we must caution the 
reader that the dynamic range of the X-ray luminosities of single stars in 
\citet{LopezMorales2007} is very small and the overall scatter of the 
X-ray values will mask any correlation. Thus, the lack of inconclusive 
evidence in the single star analysis is not surprising.

There is yet another observational piece of evidence that adds to the 
discussion on the effects of activity on stellar properties. This is the 
detailed analysis of the only known eclipsing brown dwarf reported by 
\citet{Stassun2007}. The surprising conclusion of the analysis was a 
statistically significant temperature reversal (i.e., the more massive 
brown dwarf is also the cooler of the pair), which is not compatible with 
the predictions of models. However, a recent study by \citet{Reiners2007} 
indeed shows that the more massive component is significantly more active 
than its less massive counterpart. Put in the context of our results 
above, it seems clear that higher activity would result in cooler 
temperature thus explaining the reversal found in the light curve 
analysis.

But there are also results that seem to be at odds with the hypothesis of 
stellar activity as being responsible for the $\sim$10\% differential in 
stellar radii. The recent study by \citet{Torres2007} used the 
particularities of planetary transits to determine the physical properties 
of the host star GJ~436. This star is relatively inactive, with a value 
$\log L_{\rm X}/L_{\rm bol}=-5.1$ (compared with $-3$, which is typical of 
eclipsing binaries). Surprisingly, the author also found a 10\% radius 
differential when compared with models, in agreement with the results from 
eclipsing binaries. More examples of such objects with transiting planets 
are desirable to draw any firm conclusions. However, in the meantime, we 
call attention to the fact that the analysis of GJ~436 is heavily based on 
an assumed stellar effective temperature, which is known to be poorly 
established in the low end of the main sequence, and thus potentially 
subject to large systematic errors.

\section{Starspot coverage}

One of the consequences of the starspot scenario discussed above is the 
need for a $\beta$ value of 0.3 and thus about 50\% surface coverage of 
spots cooler than the photosphere by 15\% to explain the observed radius 
discrepancies. It is worth reviewing now if such high coverage is 
compatible with the observations. Interestingly, large spot coverages are 
intuitively associated with large photometric variations. This is indeed 
not the case. Photometric variations are only sensitive to the {\em 
contrast} between different areas on the surface of the star. Thus, a 
heavily spotted star with a homogeneous spot distribution would have its 
overall light level severely diminished but display no significant 
variations along the rotation phase.

\begin{figure*}[t!]
\centering
\includegraphics[width=\textwidth]{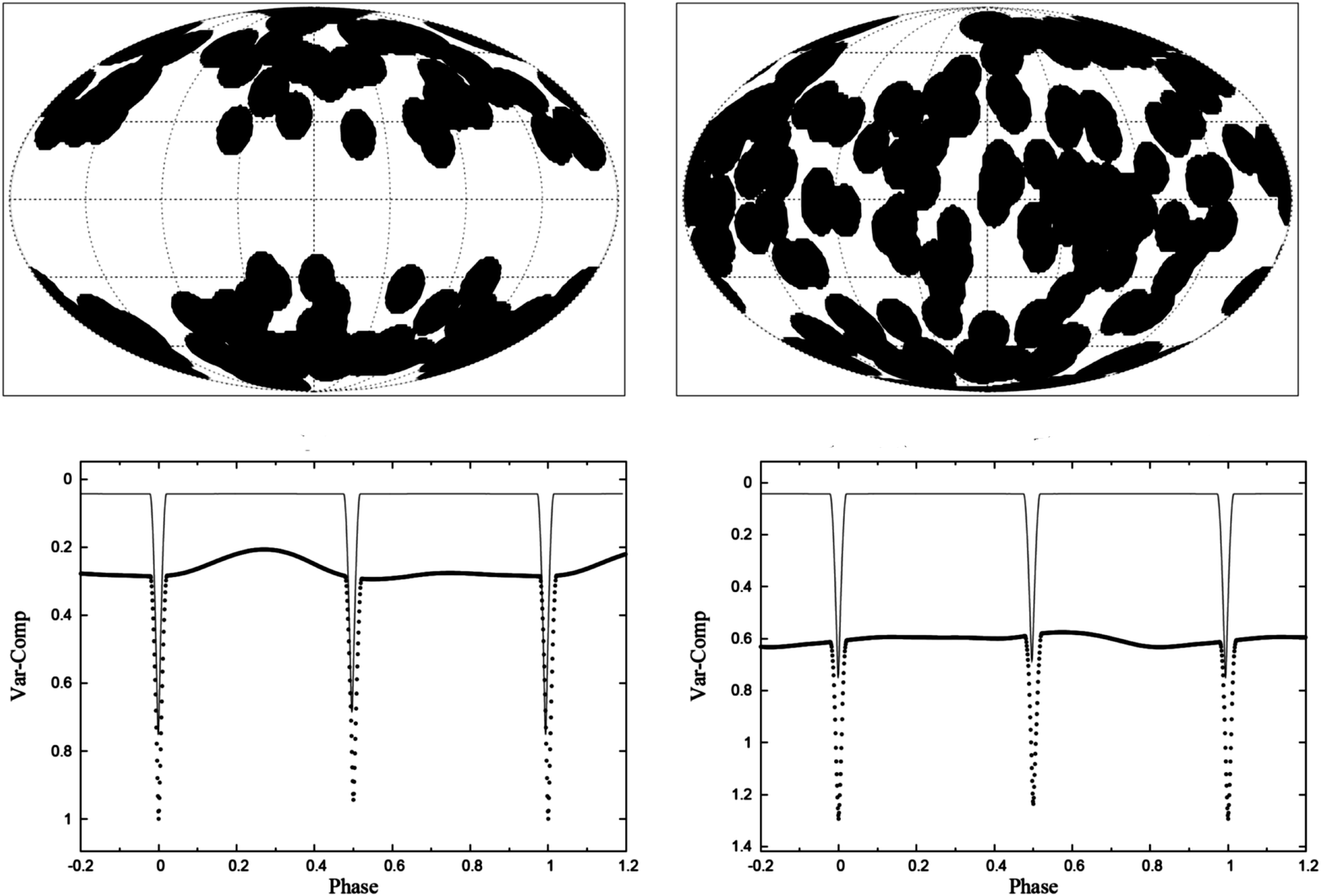}
\includegraphics[scale=0.4]{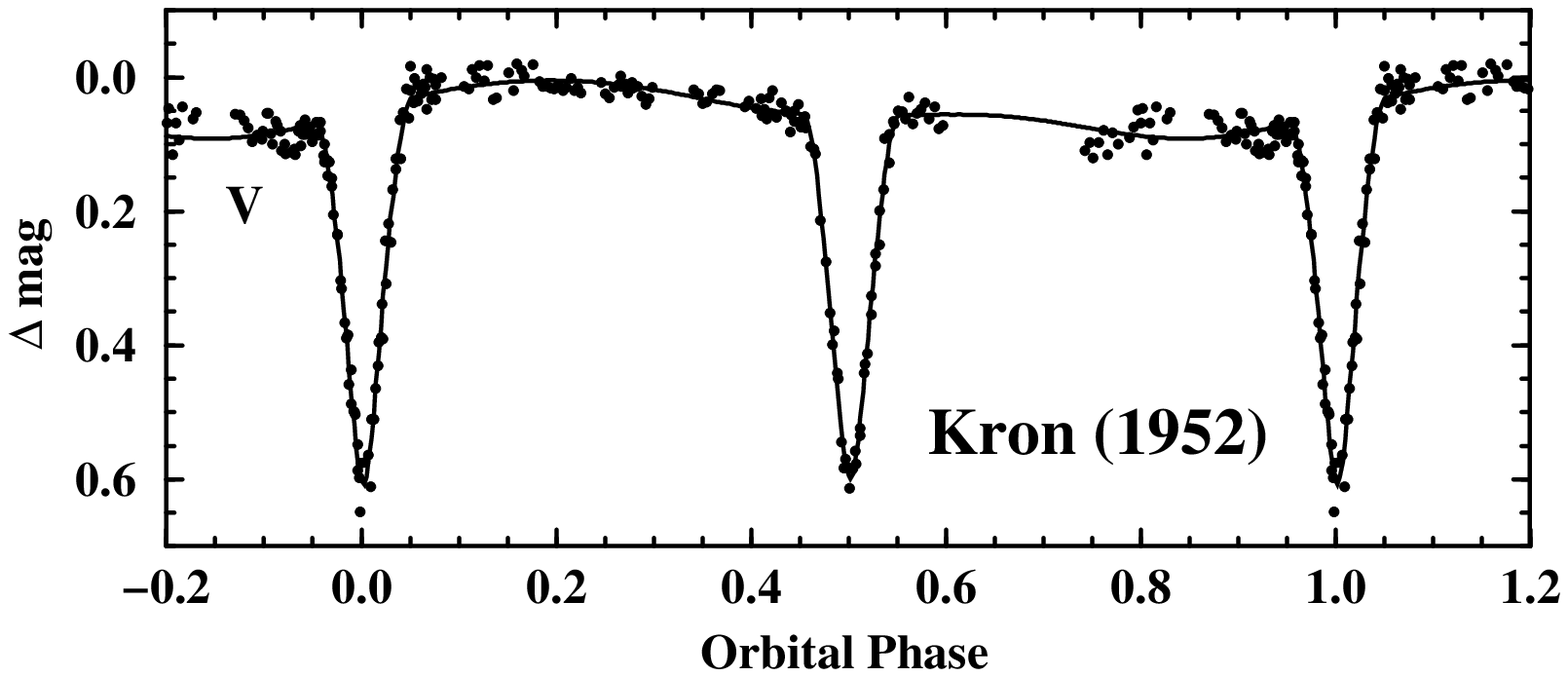}
\caption{\footnotesize
Four top panels: Plots depicting two different latitude distributions of 
starspots covering 50\% of the stellar surface. Note the different 
out-of-eclipse light drop and the resulting phase variations. Bottom 
panel: Observed light curve of the eclipsing system YY Gem in the $R$ 
band.}
\label{fig:fig_spots}
\end{figure*}

We carried out various simulations to investigate if the eclipsing binary 
data in the form of light curves are compatible with the inferred surface 
spot coverage of about 50\%. This was done by assuming different latitude 
distribution models and then evaluate the peak-to-peak magnitude 
variations. A detailed summary of our investigations will be provided 
elsewhere but here we point out that the simulated light variations out of 
the eclipses and the depth of the eclipses themselves are compatible with 
the observed light curves of several analyzed eclipsing binaries. An 
illustration of this can be seen in Fig. \ref{fig:fig_spots} for the case 
of the eclipsing binary YY Gem.

\section{Other scenarios}

Stellar activity is not the only scenario advocated to explain the 
reported differences between observed and model calculated stellar radii. 
\citet{Berger2006} analyzed a sample of stars with interferometric radii 
and found quite a strong correlation between radius differences and 
stellar metallicity. From this evidence the authors conclude that the 
mismatch between observation and theory could be due to a missing source 
of opacity in the model calculations. However, a reanalysis of an extended 
interferometric sample by \citet{LopezMorales2007} failed to identify such 
strong correlation. In addition, the interferometric radii determined by 
\citet{Segransan2003} with VLTI seem to agree well with model predictions. 
The possible relationship between radius differences and metallicity 
should be further investigated, both with additional data and with a 
statistically sound approach to evaluate the significance of the 
correlations found.

\section{Conclusions}

Without excluding at this point any other scenario, it seems clear from 
the tests discussed above that stellar activity plays a key role in 
defining the structure and radiative properties of low-mass stars. This 
could be very important in the context of young, and therefore active, 
low-mass stars, for example in clusters or star-forming regions. Failure 
to account for the effects of stellar activity can lead to strong 
potential biases in the determination of the ages of these objects and 
their ensembles. As discussed by \citet{Morales2007}, a simple calculation 
using the temperature differentials found for single active stars 
indicates that ages of young clusters determined from (active) low-mass 
stars could be systematically underestimated by about 40\%. Interestingly, 
this difference agrees well with the discrepancy between color-magnitude 
diagram ages and Li depletion boundary ages in young clusters (e.g., 
\citealt{BarradoyNavascues2004}).

\begin{acknowledgements}
The authors acknowledge support from the Spanish Ministerio de Educaci\'on 
y Ciencia through the program for Acciones Integradas HF2005-0249 and 
PNAyA grants AYA2006-15623-C02-01 and AYA2006-15623-C02-02, and from the 
French Picasso program 11412SB. 
\end{acknowledgements}

\bibliographystyle{aa}

\end{document}